# Adoption and Suitability of Software Development Methods and Practices

Sherlock A. Licorish
Department of Information Science
University of Otago
PO Box 56, Dunedin 9054, New Zealand
sherlock.licorish@otago.ac.nz

Johannes Holvitie, Sami Hyrynsalmi & Ville Leppänen
Department of Information Technology
University of Turku
FI-20014, Turku, Finland
{jjholv, sthyry, ville.leppanen}@utu.fi

Rodrigo O. Spínola & Thiago S. Mendes
Fraunhofer Project Center for Software, and
System Engineering at UFBA
Salvador University
Salvador, Brazil
rodrigo.spinola@pro.unifacs.br, thiagomendes@dcc.ufba.br

Stephen G. MacDonell & Jim Buchan
School of Computer and Mathematical Sciences
Auckland University of Technology
Private Bag 92006, Auckland 1142, New Zealand
{smacdone, jim.buchan}@aut.ac.nz

**Abstract**

In seeking to complement consultants' and tool vendors' reports, there has been an increasing academic focus on understanding the adoption and use of software development methods and practices. We surveyed practitioners working in Brazil, Finland, and New Zealand in a transnational study to contribute to these efforts. Among our findings we observed that most of the 184 practitioners in our sample focused on a small portfolio of projects that were of short duration. In addition, Scrum and Kanban were used most; however, some practitioners also used conventional methods. Coding Standards, Simple Design and Refactoring were used most by practitioners, and these practices were held to be largely suitable for project and process management. Our evidence points to the need to properly understand and support a wide range of software methods.

## I. Introduction and Background

Large bodies of work have considered the adoption and use of software development methods, more recently with a particular emphasis on Agile approaches [1]. Software development methods provide guidance for the planning and management of software projects by dividing the process into distinct phases. Earlier methodologies emphasized a planned process, where typically the output of one phase of the project is used as input for the subsequent phase in progressing the project to completion using a waterfall process [2]. Strict adherence to this approach has been found to be unsuitable for much software development, and has been blamed for numerous software project overruns and failures [3]. Such criticisms inspired new (Agile) approaches to software development that are more iterative and fluid, where software features are designed, constructed, and deployed in parallel, with a reduced emphasis on plans and processes [4]. These latter Agile approaches have received substantial attention over the last two decades. Industrial surveys [5] and those from tool vendors (e.g., VersionOne and ThoughtWorks) have also sustained this attention. These surveys have provided overwhelming support for the use of Agile methods, particularly Scrum and Extreme Programming (XP) [6], which are said to be highly relevant in contexts requiring high responsiveness to change. This support has led to the development of various tools to evaluate organizations' agility, e.g., Nokia test for Scrum[1] and How Agile Are You[2]. Furthermore, there are many agile groups that provide forums for practitioners to discuss issues specific to agile development (e.g., Agile Lean Europe, Agile Cambridge UK, Agile India).

While there has been significant *industrial* attention aimed at studying, evaluating and supporting agile methods, some believe that this could be motivated, and influenced, by self-interest [7]. Others have also expressed reservation over the lack of application of rigorous scientific methods in industry-based evaluations, and have encouraged academic-led studies as a potential means of validation [8]. We have thus looked to provide the requested robust investigation into the adoption of software development methods and practices by adhering to recommended guidelines for executing surveys

---

[1] http://agileconsortium.blogspot.co.nz/2007/12/nokia-test.html
[2] http://www.allaboutagile.com/how-agile-are-you-take-this-42-point-test/

[7]. In addition, we sought to evaluate software practitioners' perceptions of the suitability of specific software development practices. We outline four research questions to address these objectives:

**RQ1.** What are the general characteristics of contemporary software development projects?

**RQ2.** What are the characteristics of the teams and clients involved in contemporary software development projects?

**RQ3.** What software development methods and practices do practitioners adopt for their development portfolio?

**RQ4.** Are popular agile practices assessed to be suitable for software project and process management?

The first three questions (RQ1-RQ3) are designed to gather insights into the current state of practice, while the last question (RQ4) evaluates the suitability of such practices for software project and process management. Note that in this work, process management is taken to mean the management of the software development process which is considered to be repeatable across many projects (e.g., managing the quality assurance process). On the other hand, project management focuses on managing a project in the achievement of a stated outcome (e.g., delivering a financial accounting software module).

In the next section (Section II) we provide our research setting, and in Section III we present our results. We then discuss our findings and outline their implications, before considering the limitations of the work in Section IV.

## II. RESEARCH SETTING

We conducted a web-based survey across Brazil, Finland, and New Zealand to answer the research questions defined in Section I. The Google Forms platform[3] was selected to host the survey. Our instrument comprised 37 questions, including both closed- and open-ended aspects. The first 16 survey questions were closed-ended and were intended to gather data to enable us to answer the four research questions outlined in Section I[4]. The latter part of the instrument focused on the effects and management of technical debt. Findings from this second aspect of the work are to be presented in a subsequent study.

We hosted the survey separately for each country (Brazil, Finland, and New Zealand) through the use of Google Forms. Access to the survey was anonymous for respondents, and their participation was solicited from industry-affiliated member groups, mailing lists and research partners (e.g., Institute of IT Professionals New Zealand and Software Entrepreneurs Association in Finland). The instruments hosted in Brazil and New Zealand were completed only by respondents in these countries, while 95% of those responding to the Finland instrument resided there (with 5% from other European countries). The survey was hosted for three months (March 2015 to May 2015), before it was taken offline for data analysis. We have evaluated our outcomes against those reported in the VersionOne 2014 survey, asserted to be one of the most comprehensive industry-based studies [7].

In order to characterize respondents' recent projects (**RQ1**) we queried the number of software projects their host company was undertaking, the number of concurrent projects their team was responsible for, the typical duration of software projects and the number of software practitioners employed. We also sought respondents' input regarding the nature of the projects undertaken in terms of Transparency, Predictability, Efficiency, Agility, and Sustainability. Team and client characteristics (**RQ2**) were examined via team size, type of software solution, practitioners' roles, and experience. In designing this aspect of the survey we utilized a previously employed classification for software development experience [9], where the options were "under three", "three to six", and "over six" years, which provides a rough division of expertise for novice, intermediate, and experienced practitioners. We next asked practitioners to indicate which methods and practices they used for software development (**RQ3**). Options included in the survey covered both conventional (e.g., Waterfall, Spiral, Rapid Application Development) and Agile (e.g., Scrum, Crystal, Kanban) approaches. We also included specific practices that are held to be used across both conventional and agile contexts. Responses to these questions followed a five-point Likert scale – "systematic", "mostly", "sometimes", "rarely", and "never". This allowed us to correlate the use of practices with methods and other variables. Finally, we sought to obtain insights into the suitability of the practices used in relation to project and process management (**RQ4**). We asked practitioners to rate how well (or poorly) each practice covered their management needs, and their software development process.

## III. RESULTS

A total of 184 fully completed responses were gathered across the three territories: Brazil = 62 responses, Finland = 54, and New Zealand = 68. This represents an estimated response rate of 16% for Finland and 14% for New Zealand. We were not able to compute the response rate for Brazil. We now analyze these responses in answering our four research questions.

**Project Characteristics (RQ1)**: We observed that over 50% of the respondents worked in organizations responsible for between two to ten concurrent projects, while 20.7% of the firms were providing solutions to more than 25 software projects at the time. Overall, 84.2% of the respondents participated in =< 5 projects. Our results further reveal that the duration of most projects was =< 6 months. The cohort of practitioners employed tended to be variable, with 54.4% of the firms employing =< 100 employees, 9.2% employing between 101 and 250, and 30.2% employing > 250 software practitioners. Pearson Chi-Square tests were conducted to identify any statistically significant differences. The results confirm that there were significantly more firms that had a portfolio of "2 to 10" projects ($X^2 = 525.0$, df = 9, p < 0.01), significantly more teams typically handled "2 to 5" projects ($X^2 = 534.0$, df = 9, p < 0.01), and significantly more firms employed more than 250 practitioners ($X^2 = 720.0$, df = 16, p

---
[3] Refer to https://www.google.com/forms
[4] Refer to survey instrument here: http://bit.ly/agiletd

< 0.01). We observed that 87.5% of the practitioners felt that their projects were transparent, while 72.8% of the practitioners felt that their projects were *adequate*, *good* or *very good* in terms of predictability. Practitioners also highly rated the efficiency (85.3%) and agility (84.8%) of their projects.

**Teams' and Clients' Characteristics (RQ2)**: Results revealed that 89.1% of the practitioners operate in teams that comprised =< 20 members. We observed that 90.2% of the practitioners delivered both complete and partial software solutions, and responsibilities related to Research, Design, Development, Testing, and Management largely accounted for practitioners' duties (86.6%). We observed that 76.6% of the practitioners had > 6 years of experience, 48.4% developed software for external clients, 24.5% for internal clients, and 27.2% for both external and internal clients. Pearson Chi-Square tests confirmed that there were significantly more teams with sizes "2 to 5" and "6 to 10" ($X^2 = 708.0$, df = 16, p < 0.01), practitioners delivered significantly more complete than partial solutions ($X^2 = 368.0$, df = 4, p < 0.01), and there were significantly more practitioners with > 6 years of experience ($X^2 = 368.0$, df = 4 p < 0.01).

**Methods' and Practices' Characteristics (RQ3)**: We report the Methods used by practitioners in Fig 1 (a) and the Practices in Fig 1 (b). Fig 1 (a) shows that Scrum was most often utilized by respondents (71.2%), with Kanban (49.5%), Lean (39.7%) and Waterfall (35.3%) following in that order. It is important to note that practitioners were not restricted from selecting multiple methods, as we assumed it was possible for their teams to employ one or more of these approaches, or a hybrid approach, to various degrees of strictness. That said, when the question of systematic use was considered, it is noteworthy that we observed that 46 practitioners applied Scrum strictly as prescribed, 32 applied Kanban, 12 applied FDD, and 9 applied Waterfall. A Pearson Chi-Square test confirmed that Scrum was used significantly more than the other methods ($X^2 = 437.6$, df = 40, p < 0.01). Fig 1 (b) shows that most practitioners emphasized Coding Standards (81.2%), using Simple Design (74.5%), Refactoring (73.9%), Continuous Integration (73.9%) and 40-hour Week (71.2%). Furthermore, workflow practices such as the use of Iterations (84.2%), Iteration Planning (76.6%), Iteration Backlog (75.5%) and Product Backlog (76.1%) were highly utilized by practitioners. A Pearson Chi-Square test revealed that Coding Standards, Simple Design, Refactoring, Continuous Integration and 40-hour Week were used significantly more than the other practices ($X^2 = 411.7$, df = 40, p < 0.01).

**Suitability of Agile Practices (RQ4)**: Our results indicated that 84.8% of the respondents felt that the practices considered in Fig 1 supported project management *adequately*, *well* or *very well*, while 76.6% of respondents felt that the practices suitably covered software process management. We investigated the results regarding suitability for project and process management, which confirmed correlation; Spearman's rho (*r*) correlation = 0.73, p < 0.01. Thus, when practitioners felt that practices were suitable for project management they also typically indicated suitability for process management. (Note: Cohen's classification is interpreted as indicating a low correlation when $0 < r \leq 0.29$, medium when $0.30 \leq r \leq 0.49$ and high when $r \geq 0.50$ [10].) We thus tested for correlation between the use of practices and their suitability for project management; significant results (p < 0.01) are reported in Table I. Results show that although many practitioners indicated use of and satisfaction with practices, those that did not use the techniques mostly felt that they did not suitably cover their project management needs. Although not statistically significant, results also indicated that larger teams and those that stringently applied Waterfall methods deemed the practices in Fig 1 to be less suitable for adequately covering project management.

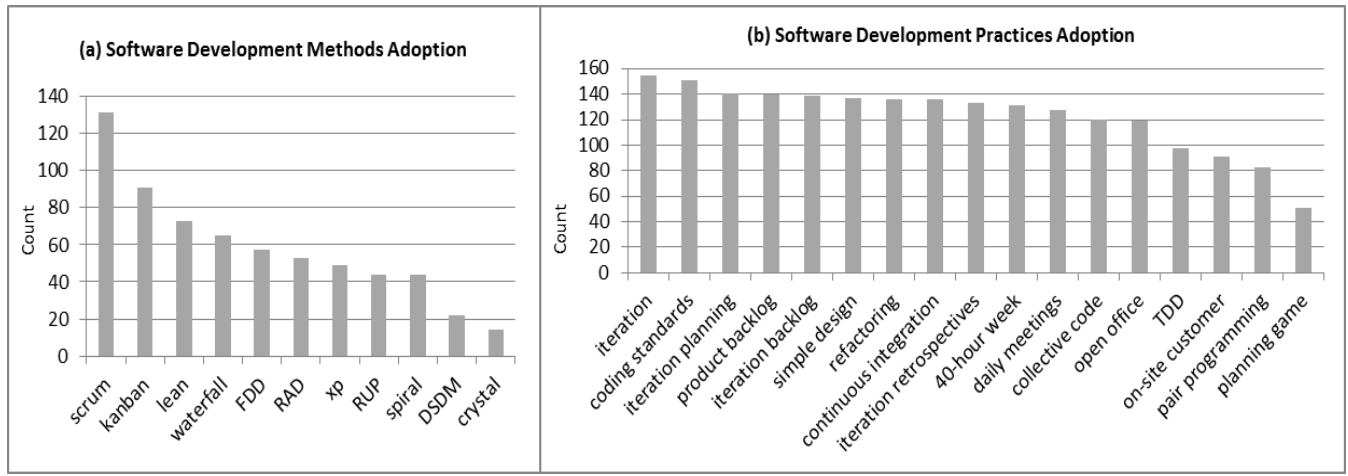

Fig. 1. Methods' and Practices' Characteristics

TABLE I.  SPEARMAN'S RANK ORDER CORRELATION RESULTS

| Software Development Practices | Correlation Coefficient ($r_s(184)$) |
|---|---|
| Simple Design | 0.37 |
| TDD | 0.45 |
| Planning Game | 0.30 |
| Coding Standards | 0.38 |
| Refactoring | 0.36 |
| Continuous Integration | 0.41 |
| Collective Code | 0.33 |
| Iteration Planning | 0.41 |
| Iterations | 0.40 |
| Iteration Backlog | 0.48 |
| Iteration Retrospective | 0.45 |
| Daily Meetings | 0.43 |
| Product Backlog | **0.55** |

**Bold** values indicate strong correlation

## IV. DISCUSSION, IMPLICATIONS AND LIMITATIONS

**Discussion and Implications**: *RQ1. What are the general characteristics of contemporary software development projects?* Our results show that practitioners developed software in organizations that were focused on a small portfolio of projects. Software projects also tended to be of mostly short duration, and 50% of the firms in our sample employed fewer than 100 employees. The results also suggested that most organizations operated in a transparent and agile manner. In comparing our outcomes to those obtained by VersionOne in 2014, we noted some disparities. For instance, we observed that 63.6% of the practitioners in our sample were working in organizations that employed fewer than 250 employees. VersionOne 2014 survey reported that 53% of the practitioners in their sample were working in organizations that employed more than 1000 employees. Of note, however, is that 65% and 21% of the respondents in VersionOne 2014 study were from North America and Europe respectively, while our respondents were split 33.7% Brazil, 29.3% Finland, and 37% New Zealand. In considering the overlap between the two surveys, we observed that respondents to VersionOne 2014 survey valued agility and predictability (similar to our findings), and felt like agile methods helped with these aspects.

*RQ2. What are the characteristics of the teams and clients involved in contemporary software development projects?* Practitioners in our sample mostly operated in small teams of "2 to 5" or "6 to 10" members, and delivered complete and partial solutions, with a predominance of Research, Design, Development, Testing and Management roles. Most respondents had > 6 years of experience, and delivered solutions for both external and internal clients. In examining our findings against those of VersionOne 2014, we observed a larger percentage of respondents who were leaders and managers for VersionOne's sample (38% against 16.6%). We observe a closer spread of development-related roles (22% against 18.8% in our study). While the VersionOne 2014 survey grouped Product Owner and Business Analysis roles, and this amounted to 13% of respondents, we noted that our Client Representative and Design roles accounted for 24%. We observed that only 28% of VersionOne 2014 respondents had > 5 years of experience, in contrast to 76.6% of our respondents.

*RQ3. What software development methods and practices do practitioners adopt for their development portfolio?* Our results show that Scrum and Kanban were used most; however, some practitioners (also) used conventional methods (Waterfall, RAD, and Spiral). Our findings are in contrast to the VersionOne 2014 outcomes, which reported that 94% of the practitioners sampled were using Agile methods. That said, VersionOne also observed Scrum use to be predominant (56% usage), and Kanban also featured prominently for the practitioners surveyed by VersionOne 2014. Earlier academic work had also replicated our pattern of results for Scrum [8]. However, these authors had noted that 18.1% of their respondents used XP. We see a divergence in the use of XP in 2015 for our study. Our results also show that Coding Standards, Simple Design, Refactoring, Continuous Integration and 40-hour Week were singled out by practitioners. We see an overlap in the responses for VersionOne's Daily Meetings, Iteration Planning and Retrospectives in Fig 1 (b), and for limited use of Planning Game and Pair Programming practices.

*RQ4. Are popular agile practices assessed to be suitable for software project and process management?* Our results indicated that common agile software practices tended to satisfy practitioners' project and process management needs. That said, practitioners that did not use these practices noted that this was due to their lack of suitability in addressing their project management needs. In particular, larger teams and those that used conventional methods did not feel that agile practices adequately addressed their needs. Of note is that the VersionOne 2014 survey, like others [1], did not examine the suitability of agile practices for project and process management.

Although not necessarily generalizable beyond our sample our findings have implications for these software engineers. In terms of practice, we note that while software practitioners may expect to work in large organizations in North America and some parts of Europe, the opposite may be more common in Brazil, Finland, and New Zealand. That said, it seems that practitioners in all locations strive for agility and predictability. Comparing outcomes in this work and those of VersionOne 2014, practitioners in Brazil, Finland, and New Zealand reported higher levels of experience than in North America and some parts of Europe. We noticed extensive use of Scrum by practitioners sampled in both studies, suggesting that practitioners skilled in the use of this method could enhance their employment prospects, regardless of location, while some practices may be location-specific (see Fig 1). Furthermore, we observe that there is ongoing relevance of conventional techniques.

**Limitations**: We considered Stavru's framework [7] in designing this work, considering various interventions to enhance the thoroughness and trustworthiness of our survey. In addition, we piloted the survey and assessed interpretations across territories, and our measures were guided by previous work. Furthermore, we tried as far as possible to target all practitioners. We also employed statistical testing to check differences in practitioners' responses across questions to enhance the work's reliability. That said, the reliability of this study could have been affected by the lack of a trained interviewer to further clarify and probe respondents' answers to the questions posed.

Finally, given space limitations, we did not test for the effects of culture and differences in everyday practices across the three locations on our outcomes, which could mediate the patterns noted. We plan to undertake this analysis in future work.